\documentclass{revtex4-1}   	
\usepackage{geometry}                		
\usepackage{graphicx}				
\usepackage{epstopdf}
\usepackage{hyperref}
\usepackage{amssymb}
\usepackage{amsmath}
\usepackage{ bbold }
\usepackage{times}

\usepackage{natbib}
\setcitestyle{round}

\newcommand{\pr}{PageRank}

\begin{document}
\title{Measuring the academic reputation through citation networks via \pr}

\author{Francesco Alessandro Massucci}
\affiliation{SIRIS Lab, Research Division of SIRIS Academic, 08003, Barcelona, Spain}
\email{francesco.massucci@sirisacademic.com}

\author{Domingo Docampo}
\affiliation{University of Vigo, atlanTTic Research Center for Communications Technologies, Campus Universitario, 36310, Vigo, Spain}
\email{ddocampo@uvigo.es}

\date{\today}							
\keywords{Page Rank, Academic Reputation, Citation Impact, Rankings, Citation Networks}

\begin{abstract}

The objective assessment of the prestige of an academic institution is a difficult and hotly debated task. In the last few years, different types of university
rankings have been proposed to quantify it, yet the debate on what rankings are {\it exactly} measuring is enduring.

To address the issue we have measured a quantitative and reliable proxy of the academic reputation of a given institution and compared our findings with well-established impact indicators and academic rankings. Specifically, we study citation patterns among universities in five different Web of Science Subject Categories and use the PageRank algorithm on the five resulting citation networks. The rationale behind our work is that scientific citations are driven by the reputation of the reference so that the PageRank algorithm is expected to yield a rank which reflects the reputation of an academic institution in a specific field.
Given the volume of the data analysed, our findings are statistically sound and less prone to bias, than, for instance, ad--hoc surveys often employed by ranking bodies in order
to attain similar outcomes. The approach proposed in our paper may contribute to enhance ranking methodologies, by reconciling the qualitative evaluation of academic prestige
with its quantitative measurements via publication impact.  
\end{abstract}

\maketitle

\section{Introduction}\label{sec:intro}
Academic institutions share fundamental missions related to the education and socialisation of students and the advancement of knowledge. Besides, there is an increasing demand on universities to make their knowledge available to society and establishing  links with the socio-economic context in which they carry out their activities. Any of these three missions, namely education, research, and social engagement, can therefore be the source of institutional reputation.

The fact of the matter is that the research mission arguably represents the most visible part of any academic outfit, and as such it has become the main source of institutional reputation. The main reason behind this is the adoption of global standards rooted on the consensus generated by peer review in the evaluation of scientific advances. Besides, research outcomes are easier to measure: there is nowadays a plethora of instruments to analyze the scientific production of individual researchers, institutions and countries. Bibliometric data, {\em i.e.} counts of papers and citations, have  been the fodder of all these instruments. Compacts of bibliometric information are periodically released by academic units and specialist consultancies \citep[e.g.][]{liucheng, waltman2012}.

Because of their relatively easier quantification, bibliometric indicators of research performance (usually related to individual and institutional research outcomes) feed a number of academic rankings \citep{Aguillo2010}.
Although it is not entirely clear to which degree the results of those rankings constitute
a trustworthy proxy of the academic reputation of a given institution \citep{Jeremic2011}, and despite scholarly agreement on the lack of appropriateness
of ranking methodologies,``rankings are now widely perceived and used as the
international measure of quality'' \citep{Hazelkorn2018}. Moreover, since rankings ``define what \textendash world-class\textendash is to the broadest audience, they cannot be ignored by anyone interested in measuring the performance of tertiary education institution'' \citep{salmi2009}.

In particular, despite its numerous weaknesses,  acknowledged willingly  by its authors, the Shanghai academic ranking of World Universities (ARWU) has  triggered reform initiatives aimed at fostering excellence and recognition, illustrating again the potency of bench-marking \citep{Aghion2008}. The results from the Shanghai ranking
have been used to assess the research strengths and shortcomings of national higher
education systems \citep{Docampo2011}, and have been shown to be reliably connected with the research excellence of a given institution \citep{Dehon2010, Docram2014}.

Our contribution is intended in the path traced by recent research in bibliometrics: just like journals evaluation cannot be captured in just one metric \citep{Moed2012}, analysing the research performance of academic institutions is also a very complex task, thus reducing it to a single measure may not be the way forward. Therefore, our investigation should be interpreted as a source of sound information to be used within or in combination with academic classification results for bench-marking purposes. In a thorough critical overview of the value and limits of world university rankings, \cite{moed2017} points to a lack of consistency in the way several well established international classifications identify scientific excellence and calls for carefully combining information from different ranking systems to get a more comprehensive view on what indicators measure. Moreover, in a recent systematic review of the usefulness of university rankings to improve research, \cite{vernon2018}  assert the need  for new measures that emphasize quality over quantity to affirm research performance improvement initiatives and outcomes. They also suggest that future research should help in evaluating three dimensions of research outcomes: scientific impact, economic outcomes, and public health impact. In line with this reflection, our contribution aims at providing new measurements that emphasise quality over quantity in relation to scientific impact and relevance. However, we would like to warn the reader again on the inherent complexity of such an endeavor, and that the results we produce should be read in combination with the results of current ranking exercises.

We begin our discussion by acknowledging that the legitimate question of
how academic rankings are effectively related with the `intrinsic' quality of a given university remains partly unanswered: indeed, one may argue that reputation is built upon the perception of excellence, which may somehow transcend mere bibliometric data. To achieve their scoring results, some rankings (e.g. ARWU) are mainly shaped by publication and staff figures, while other classifications (e.g. THE and QS) although making use of a number of bibliometric indicators rely heavily upon surveys aimed at capturing reputation. However, those surveys may be prone to biases, due to the size of the surveyed cohort, to the distribution of it across different scientific disciplines and, ultimately, to human error due to possible confusion among affiliation names.

Yet, researchers have a very concrete and measurable way to credit reputation, that is,
via citations to their peers' work. Indeed, it is very reasonable to assume that, if researcher
$x$ cites a work by researcher $y$ in one of her publications, she deems that work (and thus
its author) a reputable source of information. By means of this mechanism, it is also fair to
assume that if a researcher receives many citations, her academic reputation is globally recognised.
Finally, one may also assume that if some researcher is cited by one of her prestigious peers,
her reputation is also increased.

 This `reputation attribution' mechanism is well captured by a
well-known algorithm, {\em i.e.} the \pr~algorithm,
initially developed by \cite{pinski1976}, and later adapted  by
\cite{Brin1998}
to rank web pages according to their importance on the web.
In a nutshell, given a network of entities citing one another, the PageRank algorithm assigns a score
to each entity which is based both on the number of citations the entity receives and on the reputation of the citing institutions.
In essence, entities in the network have a high PageRank either if many other entities
cite it, or if a few other entities with a high PageRank cite it.
By aggregating the above scheme at the level of institutions, one should be able to
discern which institutions are deemed as more reputable by their peers,
by looking at how different universities cite each other.

In this paper, we specifically explore the issue of quantifying the academic reputation of a certain academic institution and to relate this measurement to the score the institution attains in a given university ranking. We do so based solely on hard bibliometric data and by exploiting the PageRank algorithm. To achieve so, we use Web of Science records on publications and citations provided by Clarivate Analytics.

Albeit one might object
 that, by analysing citation patterns, we are in fact measuring
{\it impact},
it is fair to say that \pr~applications are generally framed within the context of measuring
prestige, rather than impact  \citep[see, for instance,][]{Radicchi2011}. As we show in Sec. \ref{sec:results},
rankings based on citation counts yield in fact different results from \pr, showcasing how
\pr~is able to capture dynamics that go somehow beyond `classical' bibliometric definitions
of impact\footnote{Besides,  academic rankings, while aimed at measuring Academic Prestige,
measure, in practice, impact in some way or another -- by counting, for instance,
papers in Nature and Science, and highly cited scientists. Hence, one can fairly say that, in the
academic context at least, impact and prestige are two tightly related concepts.}. Also,
although the suitability of
the linearity of equations that lie at the core of the \pr~algorithm have been questioned by
some researchers to be correctly capturing the non-linear dynamics of scientific collaboration
and subsequent perceived prestige (see \cite{Lu2009} and \cite{Ghasemian2018}), we show in our work
that the application of \pr~on whole academic institutions, in specific academic fields,
yields very reasonable results that offer a nice compromise between academic rankings based on
bibliometric data (such as ARWU GRAS) and those largely based on reputational surveys (such as
the QS Subject Rankings).

\section{Relation to prior work} \label{sec:priorWork}
The idea of university rankings that are based exclusively on bibliometric statistics is not new. Two well-known global classifications, the Leiden ranking \citep{Leiden}, and the National Taiwan university ranking \citep{NTU} rely solely on bibliometric data. Besides, prestige-based procedures to assess scientific impact have been flourishing since \pr~was introduced in the realm of academic evaluation \citep{Zhang2017}:
\pr~has indeed been applied to rank scientific journals [in this context, see for instance \cite{Yates2015, Foulley2017},
or consider the well-known cases of the Article Influence Score \citep{Rizkallah2010}, the EigenFactor \citep{Bergstrom2007} or the Scimago Journal Ranking \citep{scimago2007}],
or to rank individual researchers \citep{Gao2016, Senanayake2015, Ying2009, Radicchi2009}. The rationale behind the use of PageRank has been well elucidated by \cite{luo2018}, by showing how prestigious citations can be  ``effectively and efficiently identified through the source affiliations of the citing paper''. \cite{nykl2014} ranked authors of scientific publications based on citation analyses, through the examination of networks of publications, authors and journals. Their results stand in support of the use of \pr~based procedures, rather than non-iterative approaches.
Also \cite{dunaiski2016} evaluated different algorithms that can be applied to bibliographic citation networks to rank scientific papers and authors. While their results recognise the relevance of citations to measure high-impact, their findings also indicate that \pr~based algorithms are better suited to rank important papers or to identify high-impact authors. \cite{Kazi2016} propose to overcome the reliance of impact methods on pure quantitative measures by using context based on three specific quality factors, namely sentiment analysis of the text surrounding the citation, self-citations, and semantic similarity between citing and cited article. Their experimental results seem to improve traditional citation counts and are similar to those rendered by PageRank based methods.

Surprisingly enough, however, \pr~has  been scantly applied at the level of academic institutions, where the
noise due to erroneous/missing publication attributions is certainly much smaller
than for the case of single researchers. Here, we make the educated guess that universities aggregate citations in the
same way as single researchers do, so that institutions with high \pr~have
a higher reputation in the network of academic institutions in a given research field.

To the best of our knowledge, the first attempt to use \pr~  based procedures to evaluate institutions was accomplished by \cite{lages2016} with the introduction of the Wikipedia ranking of world
universities: this ranking was based on the \pr~results applied to the directed networks between articles of 24 Wikipedia language editions. The Wikipedia ranking relies on a statistical evaluation of world universities which, according to their creators ``can be viewed as a new independent ranking being complementary to already existing approaches''.
\cite{lages2016} compared their \pr~list of top
100 universities with the ARWU-500 list and found a 62\% overlapping,
indicating that their analysis gives reliable results.

But Wikipedia citation patterns may escape the dynamics that actually
shape reputation in the Academic world. Therefore, we aim in this work at
applying \pr~readily to the network of citations that institutions build by citing
each others' scientific publications. The dynamics that generate those networks
should resemble more closely those that are behind the construction of academic
prestige (or the perception thereof).
Much in the same way as \cite{lages2016} used a well established global ranking to test the reliability of their results, we will also try to compare our results with well established evaluation efforts to check the validity of our approach. Clearly, citation patterns and reputation depend closely on
the academic field. For this reason, we carry out our analysis on five distinct Web of Science categories
which, incidentally, map one--to--one onto five scientific fields covered by the
ARWU thematic Global Rankings of Academic Subjects (ARWU-GRAS).
Our work helps to shed some light on how the academic prestige of ranked institutions
is captured by the GRAS rankings and
helps reconciling `qualitative' and `quantitative' ranking approaches by
providing a method that combines publication metrics and reputation in a quantitative fashion.

\section{Materials and methods}\label{sec:methods}

\subsection {A brief account of the ARWU Global Ranking of Academic Subjects\label{sec:arwugras}}
Launched in 2017, the Global Ranking of Academic Subjects (ARWU-GRAS)  ranks institutions,  presenting a minimum number of research articles in a five year period, in 52 subjects across natural sciences, engineering, life sciences, medical sciences, and social sciences. Four bibliometric indicators (related to the scientific production from 2011 to 2015 for the 2017 edition of the ranking) are present in all the subjects. For each institution and academic subject, those indicators are:
\begin{itemize}
\item[PUB] number of papers ``article'' type) authored by an institution.
\item[CNCI] Category Normalized Citation Impact of the records used to compute indicator PUB.
\item[IC] Percentage of articles with at least two different countries in the list of addresses.
\item[TOP] Number of papers published in Top Journals.
\end{itemize}
Besides, a fifth indicator, AWARD,  related to winners of specific awards applies to 30\% of the ARWU-GRAS academic subjects. Different publication thresholds and sets of indicator weights are used depending on the academic subject. The Ranking Methodology webpage\footnote{ \href{www.shanghairanking.com/Shanghairanking-Subject-Rankings/Methodology-for-ShanghaiRanking-Global-Ranking-of-Academic-Subjects-2017.html}{www.shanghairanking.com/Shanghairanking-Subject-Rankings/Methodology-for-ShanghaiRanking-Global-Ranking-of-Academic-Subjects-2017.html}} describes the indicators in more detail.
%
For each indicator, ARWU-GRAS scores are calculated following a procedure that can be summarized as follows: first multiply each value of the gathered raw data by a fixed scaling factor so that the largest raw value is scaled to 10000. Then, compress the dynamic range of the scaled raw data by taking its square root to form the indicator score \citep{Docampo2013, Docram2014}. Finally, using the weights allocated by ARWU-GRAS to each of the ranked subjects, compute the weighted sum of the indicator scores.

 The weights used by ARWU-GRAS for the indicators in the five academic subjects are listed in Table \ref{weights}.
\begin{table}[h!]
\begin{center}
\caption{   Weights allocated to ARWU-GRAS indicators in the five subjects under analysis}
\label{weights}
\begin{tabular}{lrrrrr}
    \hline
       Research Subject  & \multicolumn{1}{l}{PUB} & \multicolumn{1}{l}{CNCI} & \multicolumn{1}{l}{IC} & \multicolumn{1}{l}{TOP} & \multicolumn{1}{l}{AWD} \\
            \hline\hline
            Dentistry  \& Oral Sciences (DEN) & 100   & 100   & 20    & 100   & 100 \\
     \hline
Finance (FIN)  & 150   & 50    & 10    & 100   & 0  \\
     \hline
    Library  \& Information Science (LIB) & 150   & 50    & 10    & 100   & 0  \\
     \hline
    Telecommunication Engineering (TEL) & 100   & 100   & 20    & 100   & 0  \\
     \hline
    Veterinary Sciences (VET) & 100   & 100   & 20    & 200   & 0 \\
     \hline
    \end{tabular}%
\end{center}
\end{table}

\subsection{The data analysed}\label{ssec:data}

Five Web of Science categories have been analysed in this paper: Dentistry, Oral Surgery \& Medicine; Business, Finance; Information Science \& Library Science; Telecommunications; Veterinary Sciences. The choice of those five Categories is not accidental: indeed, they correspond to ARWU-GRAS equivalents as listed in Table \ref{weights} \citep{sarthreshold}. Moreover, three of the chosen WoS categories (Dentistry, Oral Surgery \& Medicine, Business, Finance, and Veterinary Sciences) can also be qualitatively mapped to three corresponding QS thematic rankings. QS thematic rankings are produced annually to identify top universities in a specific subjects. To do so, QS uses citations as well as global surveys of employers and academics.

The choice of scientific subjects, therefore, enables us to make meaningful comparisons between our method and well established academic rankings.
To that effect, we have downloaded the overall scores in the QS ranking of the institutions shown in their official webpage \textendash 50 institutions in Veterinary Science and Dentistry, and 200 in Accounting and Finance.
Since we can reproduce the results of ARWU-GRAS using direct data from the bibliometric suite InCites, we have made use of the official scores for institutions that are listed on the ARWU-GRAS website, and have extended the subject rankings to include all the institutions over the threshold of the minimum number of publications set by the ARWU-GRAS methodology. The list of institutions (ARWU-GRAS webpage) comprised 200 universities in all the subjects considered here, but Telecommunication Engineering (300) and Library \& Information Science (100). The total number of institutions analysed in the paper is shown in Table \ref{table:summaryTable}.
We analyse bibliometric records classified as `article-type' in the Clarivate's Web of Science database, in any of the research categories corresponding with the list of five subjects included in Table \ref{weights}. Data from the InCites platform containing all articles published in the time window 2010--2014 (both included) were provided by Clarivate Analytics in raw markup language files. For each publication, we retained the affiliation of all authors and of all references, respectively.

The data comprise 188,533 unique bibliographic records ascribed to 5,063 unique affiliations and citing 2,907,556 indexed references.
In order to compare results with the ARWU-GRAS rankings, for each WoS category only records pertaining to affiliations showing a total number of articles in excess of the publication threshold of the  corresponding ARWU-GRAS subject ranking  were retained.  In turn, only cross-citations among those
publications were considered to build the networks analysed and to compute the corresponding \pr~score.
Once this specific subset of affiliations, publications, and citations was retained for each WoS category, we ended up with
five different networks whose characteristics are reported in the next section.

\subsection{Network properties and metrics}\label{ssec:networkTheory}

In Network Theory, a weighted, directed network $\mathcal{N}({\boldsymbol n},{\boldsymbol \omega})$ is composed of
a set of $N$ nodes ${\boldsymbol n} = \{n_i\}_{i=1}^N$ and a (possibly) non--symmetric matrix of weights ${\boldsymbol \omega} = \{\omega_{ij}\}_{i,j=1}^N$.
For each $\omega_{ij}\neq 0$ it exists a weighted and directed link between nodes $i$ and $j$ of the network, with an associated weight $\omega_{ij}$.
The adjacency matrix $\boldsymbol{A}$ of the network is given by $A_{ij} = 1-\delta_{0,\omega_{ij}}$, where $\delta_{ab}$ is the Kronecker delta:
$A_{ij}$ is one whenever there is a (directed) link connecting node $i$ to node $j$ and zero otherwise.

Based on the definition above, the in--degree $k_i^{({\text in})}$ of a node $i$ is given by
\begin{equation}\label{eq:deg}
k_i^{({\text in})} = \sum_{j=1}^N A_{ji},
\end{equation}
while the in--degree distribution is given by
\begin{equation}\label{eq:degP}
P\left(k\right) = \frac{1}{N} \sum_{i = 1}^N \delta_{k_i^{({\text in})} k}.
\end{equation}
Finally, the degree centrality $c_i$ of a node $i$ is equal to
\begin{equation} \label{eq:degCentr}
c_i = k_i^{({\text in})}/(N-1)
\end{equation}
and it is upper bounded by 1 (when excluding self-links).

Since the networks we analyse differ in size, their node can attain a different maximum degree.
For this reason, to make meaningful comparison among the different networks, in Fig. \ref{fig:degreeDist} we plot the degree centrality
distribution, defined as:
\begin{equation}\label{eq:degCentrP}
P\left(c^{({\text in})}\right) = \frac{1}{N} \sum_{i = 1}^N \delta_{c_i^{({\text in})} c^{({\text in})}}
\end{equation}
and refer to degree centrality when mentioning `central' nodes in Sec. \ref{sec:results}.

\subsection{The \pr~algorithm}\label{ssec:PRexplanation}
The \pr~algorithm was devised in the 1970s by \cite{pinski1976} and then popularised
in the 1990s to rank web pages according to their `popularity':
it was in fact recast as a search--engine ranking algorithm that would prioritise pages with either many
incoming web--links or with a few incoming links from highly--ranked pages.
In other words, the algorithm assigns a high score not only to those pages that are
highly connected, but also to those ones that are linked by popular websites.

In its web application, the model behind the algorithm assumes there is a web-surfer who follows links between
web pages and who, after a series of moves, gets bored and lands on a random page.
The \pr~of a given page is therefore linked to the probability a random surfer would land to the page.
The model can therefore be seen as a Markov process whereby states are pages and the
transition probabilities are given by the links among webpages. Therefore, it is not surprising that the calculation of
the \pr~is very similar to the derivation of the Markov stationary distribution. Concretely, the equation fixing the \pr~$\boldsymbol{\pi}$ reads:
\begin{equation} \label{eq:pagerank}
\boldsymbol{\pi} =  \frac{1-d}{N} \mathbb{1} + d ~ \tilde{\boldsymbol{\omega}}\boldsymbol{\pi},
\end{equation}
where  $\mathbb{1}$ is the unitary $N$-dimensional vector and where the elements $\tilde{\omega}_{ij}$ of the matrix
$\tilde{\boldsymbol{\omega}}$ are given by $\tilde{\omega}_{ij} = \omega_{ij} \left(\sum_i \omega_{ij}\right)^{-1} $, with
${\boldsymbol \omega} = \{\omega_{ij}\}_{i,j=1}^N$ the weight matrix of the network considered (see Sec. \ref{ssec:networkTheory} above).
The quantity $d$ is called `damping factor' and is linked to the probability of leaving the current page and landing on a random website.
This factor, together with the first term on the rhs of Eq. \eqref{eq:pagerank} are
included to ensure a transition when landing on
a page without out--going links, so to preserve the
ergodicity of the process and to ensure the convergence of $\boldsymbol{\pi}$
to a unique stationary density \citep{Masuda2017}.
The \pr~is usually computed iteratively, with an initial guess $\boldsymbol{\pi}^{(1)}$ that gets updated by applying Eq. \eqref{eq:pagerank} above as:
\begin{equation} \label{eq:pagerankIter}
\boldsymbol{\pi}^{(n+1)} =  \frac{1-d}{N} \mathbb{1} + d ~ \tilde{\boldsymbol{\omega}}\boldsymbol{\pi}^{(n)}, \quad 
\boldsymbol{\pi}^{(n)} \stackrel{n \to \infty}{\to} \boldsymbol{\pi}.
\end{equation}
The result of this computation yields the $N$--dimensional $\boldsymbol{\pi}$ vector that expresses the probability of visiting
any given page $i = 1,\ldots N$, {\em i.e.}, in other words, the `popularity' of that page. In our work, we deal with bibliographic
citations among institutions in a fashion akin to links among webpages. In this setting, we effectively derive the popularity $\boldsymbol{\pi}$ of
any given institution in a network of citations.

Note that, at odds with other approaches (see, for instance, \cite{pinski1976}), we use here an
unweighted version of the \pr~algorithm, which does not take into account
the total number of publications produced by each institution. We chose this path
because the disciplinary areas we analyse are rather compact and the typology (and
size) of the different actors is fairly comparable: the size of the institutions considered here
is narrowly distributed around the mean, modulo some few extremely small
institutions that (we checked) would be over-rewarded by a normalised version of the \pr~algorithm.
For this reason, we finally decided to apply an unweighted version of the \pr~algorithm in the present work.

\section{Computing the University reputation via \pr~} \label{sec:results}
Our aim is to measure aggregate citations from a certain institution to some
other academic body, in a given field. 

 \begin{table}[h!]
\begin{center}
\caption{A summary of the data analysed. We report here, for each WoS category we analysed,
the number of unique affiliations (Unique affil.), of publications (Publications) and of cross-citations (Citations)
for the 2010-2014 timespan, after retaining only those affiliations showing a total number of articles
above the threshold of the corresponding ARWU-GRAS subject.}
\label{table:summaryTable}
\begin{tabular}{ | l | c | c | c |}
\hline
Web of Science Category & Unique affil. & Publications & Citations \\
\hline
\hline
Dentistry, Oral Surgery \& Medicine & 325 & 33,536 & 63,701\\
\hline
Business, Finance & 434 & 16,862 & 27,527\\
\hline
Information Science \& Library Science & 417 & 12,488 & 15,147\\
\hline
Telecommunications & 639 & 47,155 & 64,812\\
\hline
Veterinary Sciences & 330 & 48,074 & 50,063\\
\hline
\end{tabular} 
\end{center}
\end{table}


To that end, we aggregated all publications at the affiliation level, and were thus able to reconstruct for each WoS category the
web of cross--citations among institutions  built in the 2010--2014 time period.
The resulting system consists of a weighted network $\mathcal{N}({\boldsymbol n},{\boldsymbol \omega})$,
where each node $i \in {\boldsymbol n} $ is an academic institution
and where weighted edges $\omega_{ij} \in {\boldsymbol \omega}$ are the total number of citations occurring within the specific WoS category from
publications produced by institution $i$ to publications produced by institution $j$.

\begin{figure}[t!]
\begin{center}
\includegraphics[width=0.9\textwidth]{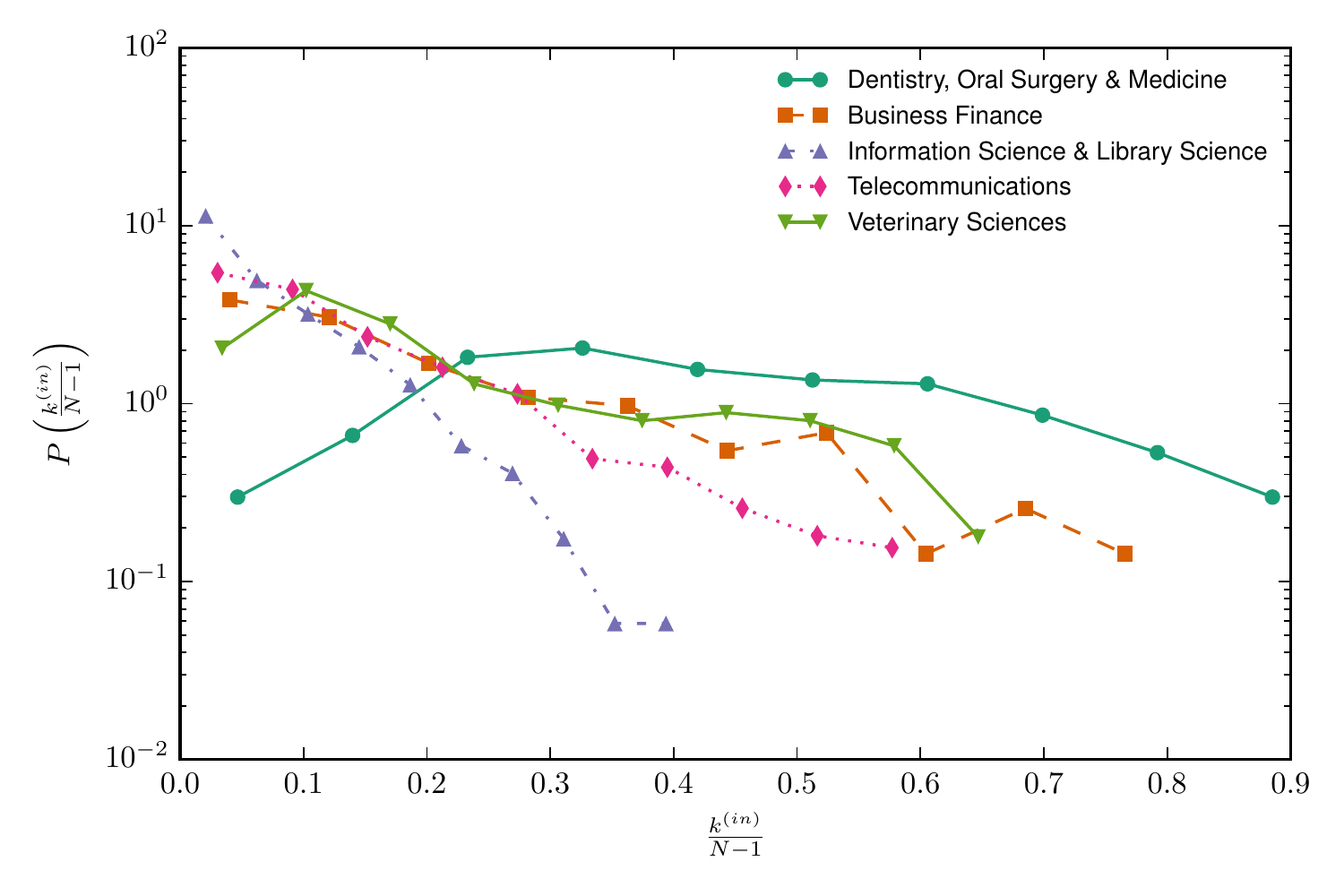}
\end{center}
\caption{The in--degree centrality $P \left(k^{(in)}/(N-1)\right)$ [Eq. \eqref{eq:degCentrP}] for the five networks considered.
These curves show the distribution of citations that institutions receive from peer academic institutions,
normalised over the size of the cohort of institutions considered for each WoS category.
It appears how Dentistry, Oral Surgery \& Medicine and Information Science\& Library Science are
the most and the least interconnected citation networks, respectively.}
\label{fig:degreeDist}
\end{figure}

The network characteristics  we obtained for each category are summarised in Table \ref{table:summaryTable},
where we report the total number of records, the total number of institutions to which the authors of
these publications were affiliated, and the total citations retrieved in the dataset. Fig. \ref{fig:degreeDist} shows
instead the in--degree centrality distribution $P(k^{(in)}/(N-1))$ of those networks (see Sec. \ref{sec:methods} for details):
these statistics allow to get a glimpse of the structure of each network and
enable one to understand how connected the hubs in network are.

For each WoS category analysed, we computed the weighted
\pr~of the resulting network of citations.
This calculation allowed us to assign a score to each academic institution: this score is related both to `quantity' through the number of aggregated citations any given institution has received and to   `quality' according to the provenance of those citations (for more details see Sec.  \ref{sec:methods}).
The rationale behind our approach is that researchers are expected to cite
the most reputable source in their publications and that entities cited by reputed sources
are expected to be, in turn, reputable. As a consequence of this mechanism, one expects
to be able to quantify academic reputation by measuring citation patterns via \pr~and without
the means, for instance, of dedicated surveys.

Note that since \pr~takes into account the reputation of citing institutions, one can
circumvent shortcomings due, for instance, to the emergence of `citation cartels' \citep{Franck1999},
which are not accounted for by mere citation counts.
The above phenomenon has been in fact observed in a few instances \citep{Fister2016}, especially in
relation to research evaluation schemes that take into account the citations researchers receive \citep{Haley2017}.
Therefore our method allows to strongly discount those effects due to the appearance of
clusters of institutions citing each other more frequently than expected.

As we discuss further below
in the following subsections, \pr~results are qualitatively in good agreement with the QS thematic rankings
for all those areas covered by QS: this first finding suggests that \pr~is indeed capable of reproducing
the academic reputation as perceived by surveys, only by means of quantitative methods.

\subsection{Viability of the \pr~algorithm to rank academic institutions}\label{ssec:}
Before getting into deeper analyses, we wanted to assess the soundness of our approach against
some well established academic ranking standard. To that aim,
 we compared the \pr~scores we obtained for each institution in each WoS category
with their respective score in the corresponding ARWU-GRAS ranking.
ARWU-GRAS rankings are just one possible benchmark of our results. We chose to
compare with them because: {\em i.} there is a precise mapping that links the different
WoS Categories to each ARWU-GRAS Subject, so that we are sure we are comparing
apples with apples, and {\em ii.} ARWU-GRAS are built on bibliometric indicators
computed from the InCites suite, so that the data consistency is ensured.

\begin{figure}
\begin{center}
\includegraphics[width=0.9\textwidth]{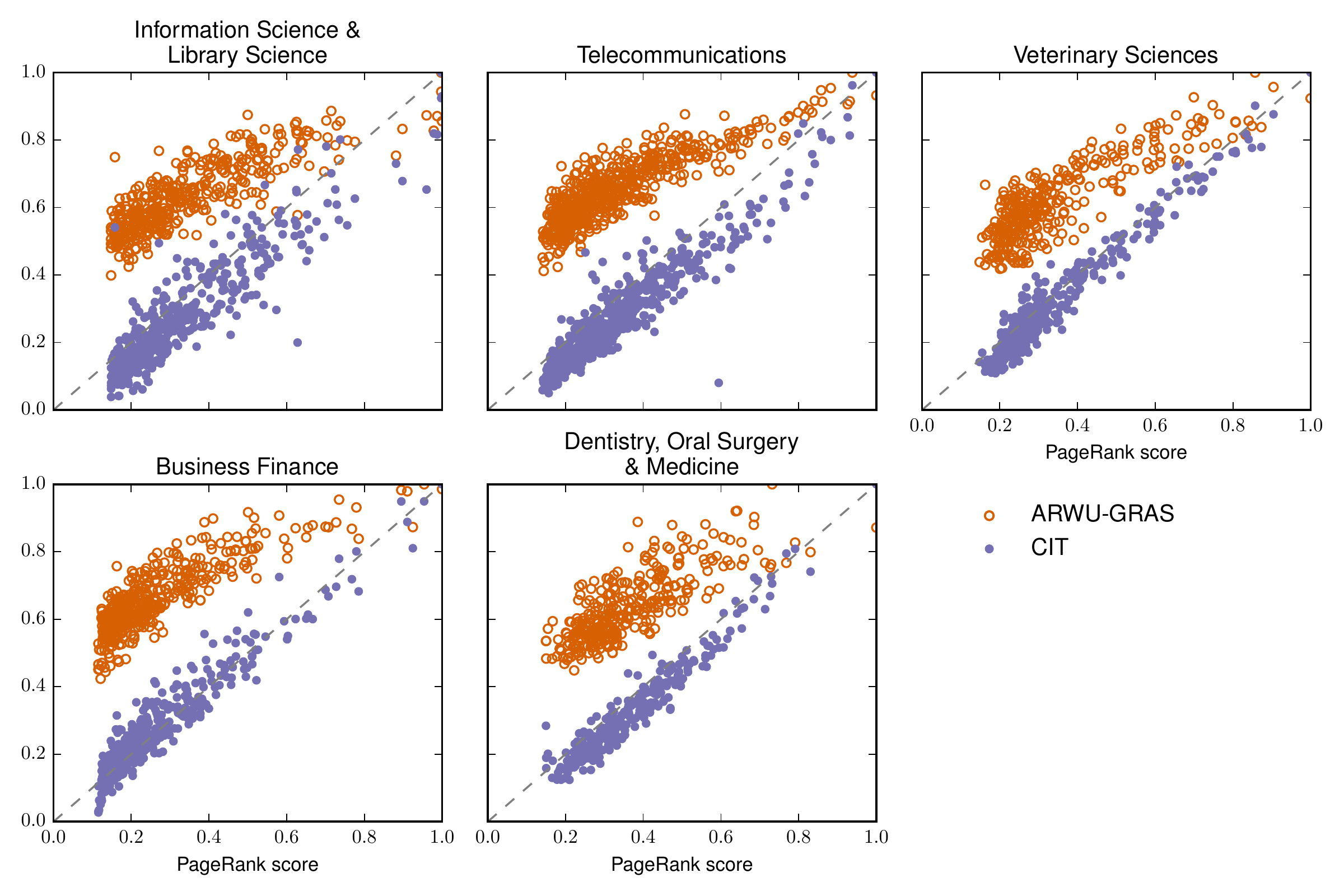}
\end{center}
\caption{Comparison of the \pr~scores against the ARWU-GRAS and citation (CIT) scores for each WoS category and institution considered in our analysis.}
\label{fig:correlations}
\end{figure}

To carry out our comparisons we started by computing the Pearson and Spearman Correlations, as well as Kendall's  coefficient of concordance between the two scoring systems.
Before computing the Pearson Correlation we proceeded to normalise the \pr~scores using the ARWU-GRASS  compression procedure explained in Section \ref{sec:arwugras}, {\em i.e.} we took the square root of the \pr~scores normalised
over the maximum score.
Table \ref{correlarwu} reports Pearson's and Spearman's correlation values, as well as Kendall's  coefficient of concordance between ARWU-GRAS and PageRank scores, while in Fig. \ref{fig:correlations} we show, for each WoS category, a scatter plot comparing the \pr~ and the ARWU-GRAS normalised scores as well as the citation (CIT) score, which, albeit not considered in ARWU-GRAS could be considered as a viable, simpler alternative to \pr. According to the results shown in Table \ref{correlarwu}, we found that  ARWU-GRAS scores and \pr~results have a significant correlation. Besides, the concordance between rankings as signaled by the non parametric statistic Kendall's W, the most familiar measure for concordance \citep{marozzi2014}, is strong across the five subjects. However, as one can observe in Fig. \ref{fig:correlations}, ARWU-GRAS (and CIT, for that matter) and \pr~ are by no means interchangeable: the point clouds are in fact rather widely scattered around a trend line at fixed PageRank values.

\begin{table}[h!]
\caption{ARWU-GRAS and PageRank scores correlation coefficients (Pearson and Spearman) and concordance coefficient (Kendall W)}
\begin{center}
\label{correlarwu}
\begin{tabular}{ | l | r | r | r |}
\hline
Research Subject 	&	Pearson	&	Spearman	&	Kendall's W	\\
\hline\hline
Dentistry, Oral Surgery \& Medicine 	&	0.79	&	0.74	&	0.88	\\
Business, Finance 	&	0.87	&	0.82	&	0.92	\\
Information Science \& Library Science 	&	0.86	&	0.85	&	0.91	\\
Telecommunications 	&	0.88	&	0.85	&	0.91	\\
Veterinary Sciences 	&	0.89	&	0.78	&	0.89	\\
\hline
\multicolumn{4}{| l |}{All Pearson's and Spearman's correlations are significant ($p<.001$)}\\
\hline
\end{tabular} 
\end{center}
\end{table}

To better understand the source of the observed correlations, and to check to what extent it is due, {\em e.g.}, to size effects,
we performed two further analyses:
\begin{enumerate}
\item we performed a Principal Component Analysis (PCA) on the overall set of the 5 WoS Categories;
\item we carried out a partial correlation analysis between the \pr~results and the ARWU-GRAS rankings.
\end{enumerate}

\begin{figure}[t!]
\begin{center}
\includegraphics[width=0.9\textwidth]{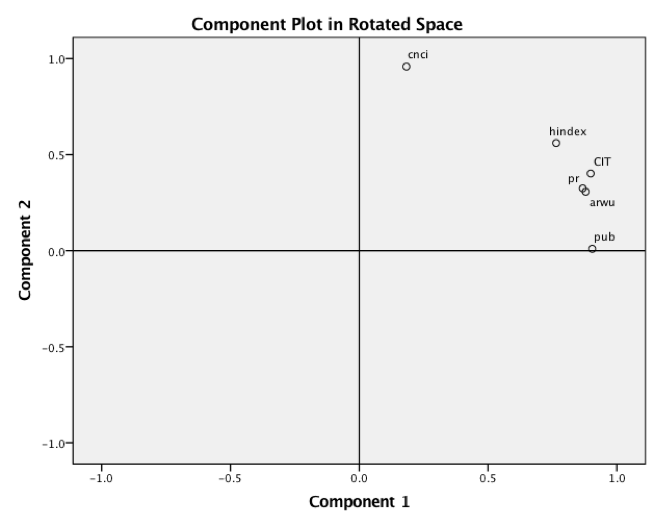}
\end{center}
\caption{A Principal Component Analysis plot that compares of a few metrics with the ARWU-GRAS scores.
This plot allows to understand which metrics ARWU-GRAS are better captured by the ranking total score
and which one resembles most the \pr~results: the ARWU-GRAS total score (score) is found to be sitting
very close to the \pr~results (pra), meaning that \pr~and ARWU-GRAS capture very similar features of
the cohort of academic institutions considered. Note that the first component of the plot is mostly aligned
with the `PUB' metric, which  is tightly correlated with the size of the institution, while the second component is mainly
aligned with the size-independent `CNCI' metric: both \pr~and ARWU-GRAS are found to be
somewhere in between these two limits.}
\label{fig:PCA}
\end{figure}

Thus, we first merged all WoS categories and performed a Principal Component Analysis (PCA)
on the space spanned  by the  following metrics: Category Normalized Citation Impact (cnci),
    Total number of publications (pub),
    Total number of citations (CIT),
    ARWU-GRAS total score (arwu),
    \pr\  score (pr), and
    H-index (hindex).
The merged dataset we used to perform PCA consists thus of 2145 observation points in a 6\textendash dimensional space, with the correlation matrix shown in Table 4. We retained two principal components which jointly contribute to explain in excess of 89\% of the variance in the sample. By projecting the original 6 dimensions onto a reduced 2\textendash dimensional space, one is
able to check which of the above metrics lie closer together (i.e. which metrics
capture similar features). The result of this effort is shown in Fig. \ref{fig:PCA}, which shows how the metrics listed above lie in the PCA reduced space after varimax rotation. The first and second rotated component explain 61\% and 26\% of the variance of the sample, respectively.

\begin{table}[t]
\label{correlationtable}
\caption{Correlation matrix corresponding to the variables used for the PC analysis on the whole set of 2145 institutions}
\begin{center}
\begin{tabular}{ | l | r | r | r | r | r | r |}
\hline
Indicators	&	arwu	&	prank	&	CIT	&	hindex	&	PUB	&	CNCI	\\
\hline\hline
arwu	&		&	0.82	&	0.89	&	0.80	&	0.77	&	0.47	\\
\hline
prank	&	0.82	&		&	0.93	&	0.85	&	0.68	&	0.42	\\
\hline
CIT	&	0.89	&	0.93	&		&	0.92	&	0.79	&	0.54	\\
\hline
hindex	&	0.80	&	0.85	&	0.92	&		&	0.65	&	0.63	\\
\hline
PUB	&	0.77	&	0.68	&	0.79	&	0.65	&		&	0.26	\\
\hline
CNCI	&	0.47	&	0.42	&	0.53	&	0.63	&	0.26	&		\\
\hline
\multicolumn{7}{| l |}{All the correlations are statistically significant ($p<.001$)}\\
\hline
\end{tabular} 
\end{center}
\end{table}

Surprisingly enough, the \pr~score and the ARWU-GRAS total
score are found to be sitting very close together, suggesting that the ARWU-GRAS ranking is
effectively capable of capturing the academic reputation of a given institution.

Once we have uncovered and weighed the correlation between ARWU-GRAS and \pr\  scores, we would like next to gauge the degree of association between the two rankings by removing a series of controlling underlying variables. To that effect, we have conducted a partial correlation analysis between ARWU-GRAS and PageRank scores, by controlling both for size-dependent  (PUB and CIT) and size-independent (CNCI) variables, respectively. The rationale behind partial correlation analysis is that the value of correlation coefficients get reduced insofar as the controlling variable exerts influence in the association between the two ranking schemes.

\begin{table}[t]
\caption{Partial correlation r-values between  ARWU-GRAS and PageRank scores for the five subjects under analysis}
\label{partial}
\begin{center}
     \begin{tabular}{|l|r|r|r|r|}
    \hline\hline   
    \multicolumn{1}{|r|}{Controlling for} & none  & CNCI  & PUB   & CIT \\
    \hline\hline
       \multicolumn{1}{|l|}{Dentistry, Oral Surgery \& Medicine } & \multicolumn{1}{r|}{0.79} & \multicolumn{1}{r|}{0.71} & \multicolumn{1}{r|}{0.42} & \multicolumn{1}{r|}{-0.01} \\
      \multicolumn{1}{|r|}{p values}  & $<.001$ & $<.001$ & $<.001$ & $0.90$ \\
    \hline\hline
    \multicolumn{1}{|l|}{Business, Finance } & \multicolumn{1}{r|}{0.87} & \multicolumn{1}{r|}{0.76} & \multicolumn{1}{r|}{0.42} & \multicolumn{1}{r|}{-0.29} \\
      \multicolumn{1}{|r|}{p values}  & $<.001$ & $<.001$ & $<.001$ & $<.001$ \\
    \hline\hline
    \multicolumn{1}{|l|}{Information Science \& Library Science } & \multicolumn{1}{r|}{0.85} & \multicolumn{1}{r|}{0.83} & \multicolumn{1}{r|}{0.52} & \multicolumn{1}{r|}{-0.07} \\
     \multicolumn{1}{|r|}{p values}  & $<.001$ & $<.001$ & $<.001$ & $0.16$\\
       \hline\hline
    \multicolumn{1}{|l|}{Telecommunications } & \multicolumn{1}{r|}{0.88} & \multicolumn{1}{r|}{0.88} & \multicolumn{1}{r|}{0.65} & \multicolumn{1}{r|}{-0.04} \\
        \multicolumn{1}{|r|}{p values} & $<.001$ & $<.001$ & $<.001$ & $0.35$ \\
    \hline\hline
    \multicolumn{1}{|l|}{Veterinary Sciences } & \multicolumn{1}{r|}{0.89} & \multicolumn{1}{r|}{0.90} & \multicolumn{1}{r|}{0.70} & \multicolumn{1}{r|}{0.02} \\
        \multicolumn{1}{|r|}{p values} & $<.001$ & $<.001$ & $<.001$ & $0.67$\\    
         \hline
             \end{tabular}
\end{center}
\end{table}

The results are shown in Table \ref{partial}, and contribute to
 shed a clear light on the relationship between ARWU-GRAS and \pr:
  \begin{itemize}
\item The indicator related to publication impact (CNCI) does not bear any noticeable influence in the relationship between ARWU-GRAS and PageRank scoring. Since CNCI is an integral part of the composed ARWU-GRAS score, this fact stands in support of the difference in which ARWU-GRAS and PageRank acknowledge reputation.
\item The indicator related to the number of publications impact (PUB) exerts a moderate  influence in the relationship between ARWU-GRAS and PageRank ranking, signaling that both ARWU-GRAS and PageRank scores are to a certain extent connected with institutional size.
\item Except for Business, Finance, the partial correlations between ARWU-GRAS and PageRank scores do not reach statistical significance when controlling for the indicator CIT. Both methods appear to use up the information provided by the size-dependent indicator related to the total number of citations.
\end{itemize}

Not surprisingly, by being \pr~built upon citation patterns, the CIT variable is the one conveying most of the correlation between  \pr~and ARWU-GRAS. However, results yielded by the two approaches differ markedly when one looks at a
finer grain
the institutional changes in positions in both classifications. Within each subject, we have computed the absolute value of the change of position of all the institutions as ranked by ARWU-GRAS and \pr. We then  collected relevant descriptive statistics of the so constructed new variables to assess the extent to which  \pr \ departs from  ARWU-GRAS in recognising academic reputation.
\begin{table}[t]
\caption{Means, standard deviations,  and  key percentile values $p(50,\;75,\;90)$ for the absolute value of the differences  in position between ARWU-GRAS and PageRank scores.}
\begin{center}
\label{diff}
    \begin{tabular}{lrrrcrr}
    Subject & \multicolumn{1}{c}{N} & \multicolumn{1}{l}{Mean} & \multicolumn{1}{l}{Std.D.} & \multicolumn{1}{l}{p50} & \multicolumn{1}{l}{p75} & \multicolumn{1}{l}{p90}  \\
    DEN & 324   & 48.5  & 39.4      & 40    & 69    & 108   \\
       \hline
    FIN & 434   & 55.8  & 48.1      & 44    & 81    & 124    \\
           \hline
    LIB & 415   & 52.2  & 43.9      & 42    & 75    & 118    \\
           \hline
    TEL & 638   & 77.3  & 70.2      & 57    & 112   & 181    \\
           \hline
    VET   & 330   & 46.3  & 44.7      & 30    & 68    & 122    \\
               \hline           \hline
    \end{tabular}%
\end{center}
\end{table}

Table \ref{diff} provides, for the five subjects under analysis,  means, standard deviations, medians, and a collection of key percentile values for the differences (absolute value) in position in ARWU-GRAS and PageRank of all the institutions. We have included  the median as well as percentiles 75 and 90 which will point to the behavior of the ranking different for the $50\%,\;25\%$ or $10\%$ of the institutions, respectively. The results shown in Table \ref{diff} reveal  large position swaps in both classifications (e.g., by computing the ratio of the the last column (P90) over the second one (N), we realize that $10\%$ of the institutions in each subject suffer a rank change of about $30\%$ of the total number of universities included in the sample), uncovering differences in both ranking methodologies in spite of the observed significant correlations between them.

These differences highlight how \pr~is a useful protocol to capture reputation rather than impact:
indeed, while citations set a common trend between ARWU-GRAS and \pr~results (hence the
observed correlations), with \pr~it does not only matter how many citation an institution aggregates, but also
the provenance of them.

A finer analysis is presented for each single WoS Category in the next few sections,
where we also show and discuss briefly the properties of each resulting citation network.


\subsection{Dentistry, Oral Surgery \& Medicine}\label{ssec:dentistry}
The institutional citation network for Dentistry, Oral Surgery \& Medicine is relatively dense,
compared with the other cases analysed further below: indeed, the in--degree centrality distribution of this particular
network has a much fatter tail than those emerging from the other categories, as shown in Fig. \ref{fig:degreeDist}.
This is also consistent with the numbers reported in Table \ref{table:summaryTable},
where the citations--to--institutions ratio is the highest.
This characteristic is intrinsically related with the citation habits
of the disciplines within the field of the Health and Medical Sciences, where papers usually include more  references  as compared to other disciplines \citep{Marx2014, Crespo2013}.

\begin{figure}[t!]
\begin{center}
\includegraphics[width=\textwidth]{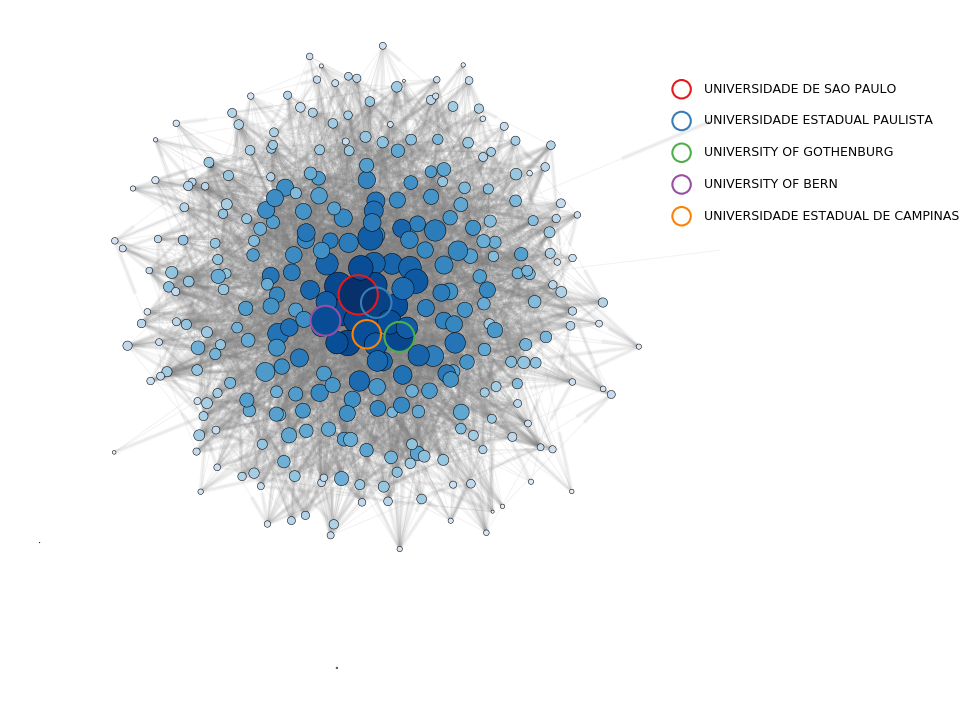}
\end{center}
\caption{The institutional network of cross-citations in the Dentistry, Oral Surgery \& Medicine WoS category.
 Edges are citations
from a publication produced by an institution to those authored by another one (10\% of the total edges are plotted).
The node size is proportional to the number of publications.}
\label{fig:dentistryNet}
\end{figure}

This feature makes, in turn,
the institutional network more interconnected, since more references per publication
directly imply more affiliations cited. This is clearly visible by looking at the actual shape
of the network analysed in Fig. \ref{fig:dentistryNet}: the network edges
({\em i.e.} citations from one institution to another) cover indeed the whole background space.

Despite the overall large connectivity, a cluster of central hub institutions (in terms of in--degree centrality) can
be easily detected in the network: these are the institutions receiving more citations
from their peers and are the ones for which we show names in Fig \ref{fig:dentistryNet}.

\begin{table}
\caption{The Top 10 \pr~institutions in  Dentistry, Oral Surgery \& Medicine.
According to the \pr~metrics and definitions, these are the most `reputable' academic institutions in this WoS category.
The three columns show the position in the \pr, the classification in the ARWU-GRAS (computed according to \cite{Docampo2013}) and the academic institution, respectively.
}
\begin{center}
\label{table:topPRdentistry}
\begin{tabular}{| c | c | l  |}
\hline
PageRank & ARWU--GRAS rank & Institution\\
\hline
\hline
1	&	9	&	University of Sao Paulo\\ \hline
2	&	26	&	University of Gothenburg\\ \hline
3	&	17	&	University of Bern\\ \hline
4	&	40	&	UNESP\\ \hline
5	&	1	&	University of Michigan-Ann Arbor\\ \hline
6	&	44	&	University of Campinas\\ \hline
7	&	50	&	The University of Hong Kong\\ \hline
8	&	41	&	Academic Center for Dentistry Amsterdam\\ \hline
9	&	25	&	University of Zurich\\ \hline
10	&	4	&	Harvard University\\ \hline\hline
\end{tabular}
\end{center}
\end{table}

When computing the \pr, we find as the best scored the University of S\~ao Paulo,
the University of Gothenburg and the University of Bern, respectively. These  institutions
are also among the central core of the citation network, meaning that those universities
not only receive many citations, but they also do so from equally prestigious institutions.
The list of the top 10 institutions for reputation measured via \pr~is given in Table \ref{table:topPRdentistry}, while
the full list is provided in the Supplementary material.
Importantly, out of these ten institutions, four of them (the University of Michigan, the University of Hong Kong,
the Academic Center for Dentistry Amsterdam and Harvard University, respectively)
are featured in the top 10 universities of the QS thematic ranking for Dentistry, in terms of the metric {\em academic reputation}.

One can also compute the Spearman's correlation coefficient on total scores between QS and ARWU GRAS and \pr, for the
the top 20 institutions of the QS Subject ranking. By doing so, one sees that the Spearman coefficient
between QS and ARWU is $\rho=0.06$, while between QS and \pr~is $\rho=0.28$.
These facts suggest \pr~is actually capable of capturing the reputation of a given institution,
as expected, and to go beyond ARWU GRAS results.
But while the academic reputation score in the QS ranking is obtained
by means of surveys, whose control in terms of significance and robustness
is hard to attain, here we derived a similar score based only on bibliometric data.

\subsection{Business, Finance}\label{ssec:finance}
\begin{figure}[t!]
\begin{center}
\includegraphics[width=\textwidth]{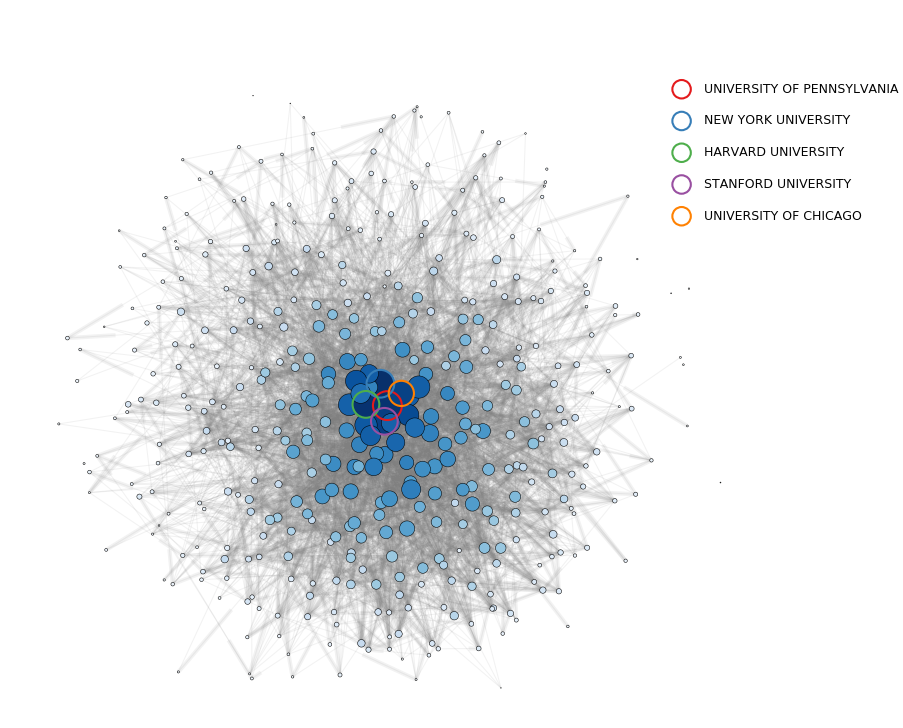}
\end{center}
\caption{The institutional network of cross-citations in the Business, Finance WoS category.
 Edges are citations
from a publication produced by an institution to those authored by another one (10\% of the total edges are plotted).
The node size is proportional to the number of publications.}
\label{fig:financeNet}
\end{figure}

The second category we study is Business, Finance. This network is the second smallest in terms of
publications and citations, suggesting there exist a fragmentation pattern in knowledge communication
and sharing in this field. By looking at the actual
shape of the network in Fig. \ref{fig:financeNet}, one can appreciate that `peripheral' nodes are
effectively little connected but that, at the same time, there is a densely inter--connected cluster
of hub institutions at the center.
The in--degree centrality distribution shown in Fig. \ref{fig:degreeDist} has indeed a fatter tail distribution in this case
if compared with the  Information Science \& Library Science WoS category, for instance.

\begin{table}
\caption{The Top 10 \pr~institutions in  Business, Finance.
According to the \pr~metrics and definitions, these are the most `reputable' academic institutions in this WoS category.
The three columns show the position in the \pr, the classification in the ARWU-GRAS (computed according to \cite{Docampo2013}) and the academic institution, respectively.
}
\begin{center}
\label{table:topPRfinance}
\begin{tabular}{| c | c | l |}
\hline
PageRank & ARWU--GRAS rank & Institution\\
\hline
\hline
1	&	2	&	University of Pennsylvania	\\ \hline
2	&	1	&	New York University	\\ \hline
3	&	15	&	Stanford University	\\ \hline
4	&	4	&	Harvard University	\\ \hline
5	&	3	&	University of Chicago	\\ \hline
6	&	29	&	Northwestern University	\\ \hline
7	&	6	&	Massachusetts Institute of Technology (MIT)	\\ \hline
8	&	20	&	Duke University	\\ \hline
9	&	5	&	Columbia University	\\ \hline
10	&	12	&	University of Michigan-Ann Arbor	\\ \hline
\end{tabular}
\end{center}
\end{table}

The \pr~results are, once again, fairly interesting. In the top three position, we find
the University of Pennsylvania, the New York University and Stanford University, respectively.
The three of them belong to the central cluster of knowledge hub institutions.
The list of top 10 institutions is given in Table \ref{table:topPRfinance}, while the full rank
is given in the Supplementary Materials. In the top 10 we find 6 institutions
(Harvard University, and Stanford University, Massachusetts Institute of Technology,
University of Chicago, University of Pennsylvania, and New York University, respectively)
that are featured in the first 10 positions in the QS {\em Accounting and Finance} thematic ranking, for {\em academic reputation}.
Also, considering the top 20 institutions featured in the QS Subject ranking, the Spearman's correlation coefficient
between the QS scores and ARWU GRAS scores is $\rho=0.59$, while for the case of \pr~it equals $\rho=0.6$.
Again, these findings are a strong indication that \pr~is indeed capable of capturing the reputation of
an Academic institution based solely on bibliometric data and to go beyond ARWU results.

\subsection{ Information Science \& Library Science}\label{ssec:library}

\begin{figure}[t!]
\begin{center}
\includegraphics[width=\textwidth]{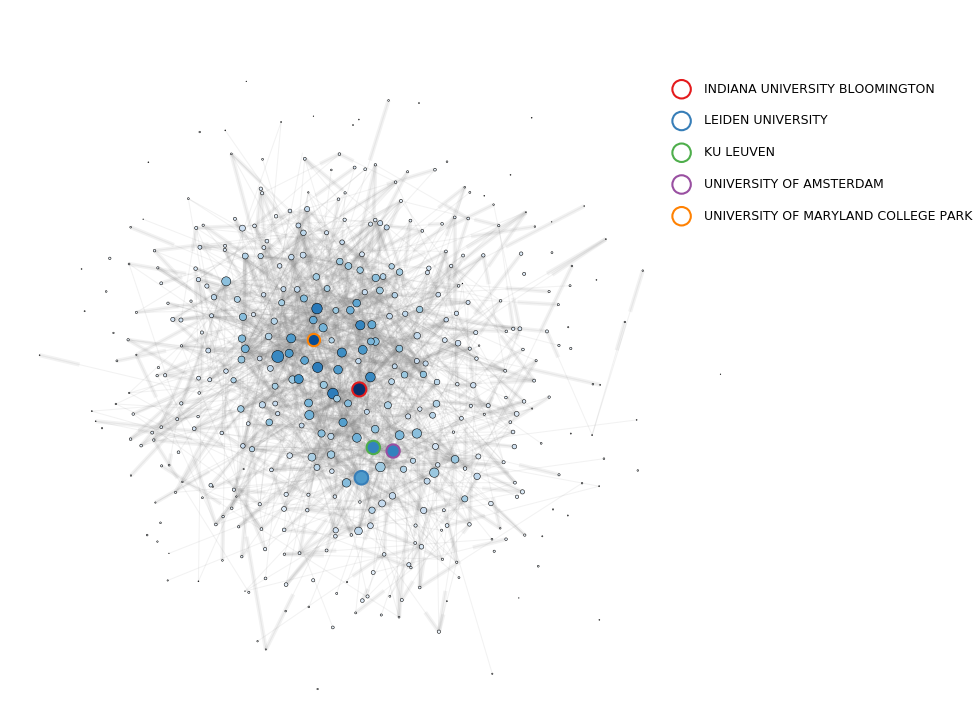}
\end{center}
\caption{The institutional network of cross-citations in the  Information Science \& Library Science WoS category.
 Edges are citations
from a publication produced by an institution to those authored by another one (10\% of the total edges are plotted).
The node size is proportional to the number of publications.
}
\label{fig:libraryNet}
\end{figure}

We next examined the case of  Information Science \& Library Science, which is
of special interest for us, considering the focus of the present work.
This field is the one with the fewer publications and citations with respect
to the set of WoS Categories analysed in this paper. The corresponding network
shown in Fig. \ref{fig:libraryNet} is indeed much sparser than the previous ones and the
respective in--degree centrality distribution decays much faster than in the other cases
(see Fig. \ref{fig:degreeDist}).

In this case, there is no sharp cluster of central hubs of knowledge as in
the previous cases, albeit some key nodes (in terms of in--degree centrality)
can be identified in the core of the
network (as, {\em e.g.}, the Indiana University at Bloomington, the University of Amsterdam and the National
University of Singapore). However, these are not as interconnected as
the central core nodes in the Dentistry case, for instance. This finding suggests that
the research communities in the field of  Information Science \& Library Science are
much more fragmented than in the other fields of research considered in the present study.

\begin{table}
\caption{The Top 10 \pr~institutions in   Information Science \& Library Science.
According to the \pr~metrics and definitions, these are the most `reputable' academic institutions in this WoS category.
The three columns show the position in the \pr, the classification in the ARWU-GRAS (computed according to \cite{Docampo2013}) and the academic institution, respectively.
}
\begin{center}
\label{table:topPRlibrary}
\begin{tabular}{| c | c | l |}
\hline
PageRank & ARWU--GRAS rank & Institution\\
\hline
\hline
1	&	8	&	University of Maryland, College Park	\\ \hline
2	&	1	&	Harvard University	\\ \hline
3	&	2	&	Indiana University Bloomington	\\ \hline
4	&	6	&	University of Amsterdam	\\ \hline
5	&	17	&	Leiden University	\\ \hline
6	&	5	&	KU Leuven	\\ \hline
7	&	14	&	University of Washington	\\ \hline
8	&	56	&	Temple University	\\ \hline
9	&	34	&	University of British Columbia	\\ \hline
10	&	32	&	University of Pittsburgh, Pittsburgh Campus	\\ \hline\hline
\end{tabular}
\end{center}
\end{table}

The \pr~analysis in this particular network yields the results shown in Table \ref{table:topPRlibrary}.
We find the University of Maryland (College Park), Harvard University and the Indiana University
at Bloomington, respectively, to be the three most reputable institutions in terms
of citations measured via \pr. The full top 10 rank of this category is reported in Table \ref{table:topPRlibrary},
while the full list is provided in the Supplementary Materials.

In this case, there is no specific QS thematic ranking to compare with, but the results
we obtain seem to be very meaningful without much further checking: we find indeed all
the `usual suspects' to be in the top 10 of the ranking. It is noteworthy the compactness
of the northern European group, University of Amsterdam--Leiden--KU Leuven: these universities
are clearly a world reference and thus high in the \pr~because of their global impact.
However, by frequently cross--citing each other, they produce a supplementary `catalysing effect'
which boosts their score towards the top of the ranking.

\subsection{Telecommunications}\label{ssec:teleco}
The next case study is the Telecommunications WoS category.
This particular network is densely connected and it is hard to detect a
central cluster of emerging institutions. A clear, emerging feature is, in this case,
the massive presence of Chinese institutions. The top 3 institutions according to the \pr~metrics are
the University of Texas at Austin, Tsinghua University and the University of Waterloo, respectively.
It is worth noting in this case that not all the top \pr~institutions are the ones with the largest
in--degree centrality: indeed,
a large number of citations does not necessarily imply a high \pr,
meaning that the reputation of the citing institutions  and not only the quantity of citations matters in \pr.

\begin{figure}[t!]
\begin{center}
\includegraphics[width=\textwidth]{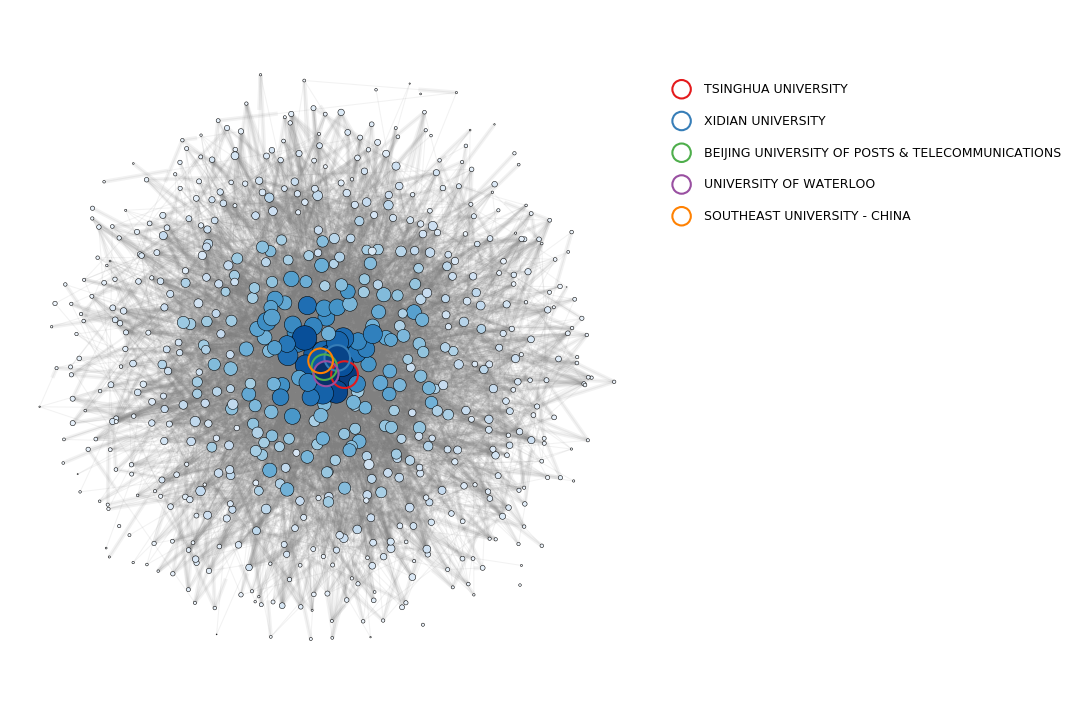}
\end{center}
\caption{The institutional network of cross-citations in  Telecommunications.
 Edges are citations
from a publication produced by an institution to those authored by another one (10\% of the total edges are plotted).
The node size is proportional to the number of publications.
}
\label{fig:telecoNet}
\end{figure}

Albeit there is no QS reputation specific ranking corresponding to this category, the top institutions we found are all
featured in the top positions of the ARWU-GRAS ranking of Telecommunication Engineering. The list
of the top 10 institutions is given Table \ref{table:topPRtelecommunication}, while the full results are
provided in the Supplementary Materials.
%
\begin{table}
\caption{The Top 10 \pr~institutions in the Telecommunication Engineering WoS category.
According to the \pr~metrics and definitions, these are the most `reputable' academic institutions in this WoS category.
The three columns show the position in the \pr, the classification in the ARWU-GRAS (computed according to \cite{Docampo2013}) and the academic institution, respectively.
}
\begin{center}
\label{table:topPRtelecommunication}
\begin{tabular}{| c | c | l |}
\hline
PageRank & ARWU--GRAS rank & Institution\\
\hline
\hline
1	&	4	&	The University of Texas at Austin	\\ \hline
2	&	1	&	Tsinghua University	\\ \hline
3	&	6	&	University of Waterloo	\\ \hline
4	&	8	&	Georgia Institute of Technology	\\ \hline
5	&	2	&	Beijing University of Posts and Telecommunications	\\ \hline
6	&	7	&	Xidian University	\\ \hline
7	&	3	&	Nanyang Technological University	\\ \hline
8	&	5	&	University of British Columbia	\\ \hline
9	&	12	&	Southeast University	\\ \hline
10	&	10	&	Shanghai Jiao Tong University	\\
\hline\hline
\end{tabular}
\end{center}
\end{table}

\subsection{Veterinary Sciences}\label{ssec:veterinary}
The last WoS category we analyse is Veterinary Sciences.
Again, the resulting network is fairly connected: in this case as well it is possible to
detect a central cluster of `influencing' institutions, that -- interestingly -- belong to very different
geographical regions. Among those, we may name for instance the University of California at Davis,
 the Royal Veterinary College and Ghent University.

\begin{figure}[t]
\begin{center}
\includegraphics[width=\textwidth]{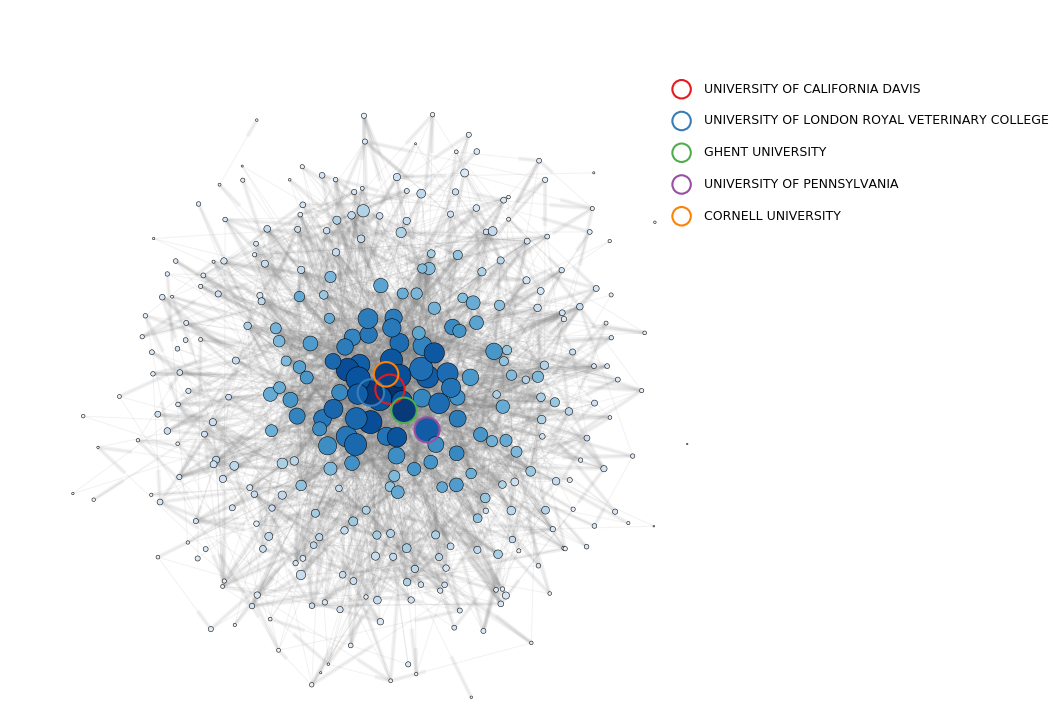}
\end{center}
\caption{The institutional network of cross-citations in the Veterinary Sciences WoS category.
 Edges are citations
from a publication produced by an institution to those authored by another one (10\% of the total edges are plotted).
The node size is proportional to the number of publications.
}
\label{fig:veterinaryNet}
\end{figure}

We computed the \pr~for the last time on this particular network and found the top three universities to be
the University of California at Davis, the Royal Veterinary College and the University of Pennsylvania:
therefore, some of the main hubs in this case do also correspond with the most reputable institutions in
the field. The top 10 \pr~institutions in this category are listed in Table \ref{table:topPRveterinary}, while the full
list is provided in the Supplementary Materials.

%
\begin{table}
\caption{The Top 10 \pr~institutions in the Veterinary Sciences WoS category.
According to the \pr~metrics and definitions, these are the most `reputable' academic institutions in this WoS category.
The three columns show the position in the \pr, the classification in the ARWU-GRAS (computed according to \cite{Docampo2013}) and the academic institution, respectively.
}
\begin{center}
\label{table:topPRveterinary}
\begin{tabular}{| c | c | l |}
\hline
PageRank & ARWU--GRAS rank & Institution \\
\hline
\hline
1	&	5	&	University of California, Davis	\\ \hline
2	&	3	&	The Royal Veterinary College	\\ \hline
3	&	19	&	University of Pennsylvania	\\ \hline
4	&	1	&	Ghent University	\\ \hline
5	&	18	&	Cornell University	\\ \hline
6	&	10	&	Colorado State University	\\ \hline
7	&	26	&	University of Guelph	\\ \hline
8	&	14	&	University of Copenhagen	\\ \hline
9	&	13	&	Swedish University of Agricultural Sciences	\\ \hline
10	&	29	&	North Carolina State University - Raleigh	\\ \hline
\hline
\end{tabular}
\end{center}
\end{table}

We can again compare the results we obtain for the top listed institutions with the QS thematic
ranking in Veterinary. Five of the institutions listed in Table \ref{table:topPRveterinary} are found to be in the top 10 QS Veterinary
when ranked for academic reputation (in particular, the first two are found to coincide), a fact that hints again
that \pr~is indeed capable of capturing the reputation of a certain institution. Furthermore,
when computing the Spearman's correlation coefficient $\rho$ on total scores between QS and ARWU GRAS and \pr, for the
the top 20 institutions of the QS Subject ranking, one finds
$\rho=0.16$ between QS and ARWU and $\rho=0.44$ between QS and \pr, again implying that
\pr~results enables one to gain more information than merely using ARWU GRAS rankings.

\section{Discussion}\label{sec:discussion}
We began this paper by recognising the intrinsic difficulty in measuring academic reputation, either through carefully designed surveys or the use of proxy indicators of performance. In spite of the complexity of academic institutions and their varied missions, we came to stress the pre-eminence of research performance as the main measurable source of academic reputation, due to the fact that scientific progress is nowadays judged through well-established peer review processes and research outcomes are easily measured in terms of numbers of publications and citations in the scientific literature.

We have then discussed the relationship between reputation and the emergence of academic rankings. Academic classifications come in different flavours, but, as we argue, they can be reduced to two main categories according to the role they assign to quantitative data. We have acknowledged two well-established worldwide rankings, THE and QS, which allocate a considerable weight to reputation scores based on academic surveys, and the ARWU-GRAS ranking, driven by publication and staff figures.

Academic rankings try to gauge institutional quality and in so doing become academic performance referees. However, one should never forget that measurements in the social sciences are a tricky business. In this regard, Adcock and Collier have sagely recommended to try and draw a clear distinction between measurement issues and disputes about concepts \citep{adcock2002}. International classifications arguably raise controversy about both sides of the issue,
that is to say, about whether the used indicators constitute valid measurements of academic performance, or whether the reputation they claim to recognise is tantamount to university excellence.
Ultimately, even though reputation is built upon the  perception of quality, which should,  to a reasonable extent, be captured by THE and QS, it is not beyond any reasonable doubt that their surveys are not  prone to biases, not the least of them being the difficulty for newcomers to play the reputation game.

Yet, we argue in the paper that researchers have put in place a solid and measurable way to credit reputation through citations to the published work of other researchers.  Hence, we reckon that there is a wealth of information at our disposal to assess reputation levels using solid data, thus dodging the biases associated with any, no matter how carefully and thoroughly designed, survey.

We acknowledge that a good `reputation attribution' mechanism has been successfully embedded in the \pr~algorithm, which has been widely popularised as a ranking tool for the relevance of web pages. In our study, we substitute  citations  among universities for links to web pages.
The rationale behind our approach is that scientific citations are driven by the quality of the cited work, and that, in turn,
citations become the more relevant the higher the reputation of the citing institution. To put it plainly, researchers are expected to cite
reputable sources in their publications; in turn, institutions cited by reputed sources
are expected to be reputable.  The \pr~algorithm  handles the trade-off between quantity of citations and
relevance of their sources and thus becomes a reliable instrument to assess the reputation of an academic center in a specific field.
For this project, we have computed the results of the \pr~algorithm within a network of
institutions citing one another
through scientific papers published between 2010 and 2014;
as a result, we have been able to render a picture of a  research subject in which institutions become ranked by reputation through analyzing  the number and provenance of their citations.

We built institutional citation networks for five WoS categories: networks are shown in Figs. \ref{fig:dentistryNet}--\ref{fig:veterinaryNet},
where each node represent an academic institution and edges citations from one institution to another. For each network, we uncovered
different structural properties (as summarised, for instance, in Fig. \ref{fig:degreeDist}), due to different citing habits in the different
research fields considered. Additionally, by using standard network metrics, we were able to detect particularly central institutional nodes
in each research field network.

We assessed the soundness of our analyses by comparing \pr~results to a well established academic ranking standard: because we have selected for our analysis five WoS categories each matching exactly one ARWU-GRAS, we have been able to relate our measurements of academic reputation via \pr~to the score the same institutions attain in the equivalent ARWU-GRAS ranking. Moreover, since three of the subjects selected also correspond to three Thematic QS subjects, we have also been able to analyze the level of association of reputation via QS surveys with measures taken using the \pr~algorithm.

We found a statistically significant association between the results rendered by the \pr~algorithm and the Shanghai Subject Rankings corresponding to the the five WoS categories analysed in the paper. The correlation between both measurements was found to be very solid, as Table \ref{correlarwu} shows. We also computed the correlation of ARWU-GRAS and PageRank results with four widely used bibliometric indicators, namely the number of publications, the number of citations, the H-index, and the category normalized citation impact to have a calibration measure to weigh the significance of \pr~ and ARWU-GRAS correlation. The general agreement of PageRank scores with ARWU-GRAS scores and the four bibliometric indicators stands in support of the reliability and validity of our results.

Furthermore, we explored the locus of the \pr~indicator in relation to the ARWU-GRAS score, when the two measures were part of a multi-dimensional space along with the four measures included with ARWU-GRAS and PageRank in \ref{correlationtable} that bear a close connection with quantity and quality of publications: from raw quantity in number of publications (PUB) and citations (CIT), to impact (CNCI), through a well-known measure that effectively combines quantity and quality, the H-index (see Table \ref{correlationtable}).
The four measures (PUB, CIT, CNCI, H-index) have been computed for the set of articles published by each institution between 2010 and 2014, within the subject under analysis. Citations were counted up until October 2017.
The merged dataset contained 2145 observations. By reducing the dimensionality using  principal components we were able to elucidate the answer to the quest for the locus of the \pr~results, since two principal components, which jointly contribute to explain in excess of 89\% of the variance of the sample, were enough to show the metrics that capture similar features. As  shown in Figure \ref{fig:PCA}, the position of the  \pr~score and the ARWU-GRAS total score in the two-dimensional principal components offer compelling evidence of the level of concordance of ARWU-GRAS and PageRank.  Moreover, in the three WoS categories selected for the analysis with equivalent subject QS rankings, we found that the results from the \pr~algorithm were in fact very well aligned with the perceived reputation as gauged by the QS surveys.

As we have pointed out in Sec. \ref{sec:results} through partial correlation analyses, the association between ARWU-GRAS and \pr~scores is mainly due to the size-dependent variable controlling for the total number of citations. However, we showed that actual ranking results vary deeply when ranking institutions for citations or via \pr. This is because \pr~accounts both the number of citations to score institutions (hence the correlation) and the provenance of them (hence the observed difference). This finding stands in support of the adoption of \pr~as a protocol for measuring reputation: indeed, an institution is reputable not only if it is highly cited, but also if it is cited by reputable institutions. In other words, with \pr, it is not only important how many people cite my work, but also {\em who} these people are.

By pointing to the similarities between ARWU-GRAS, QS and PageRank, we meant to demonstrate the reliability and validity of our approach. However, the magnitude of the correlation between ARWU-GRAS and PageRank does not mean that their scores ``convey the same information, and thus can be used interchangeably'' \citep{west2010}: to show the degree of departure of a PageRank driven classification from the one provided by ARWU-GRAS we analysed the institutional changes in positions in both rankings. After computing the changes  (absolute value), the descriptive statistics shown in Table \ref{diff} uncovered substantial differences in the way ARWU-GRAS and PageRank acknowledge reputation.

\section{Conclusion and Further Work}\label{sec:conclusion}
In this paper, we have  explored the use of a quantitive, and reliable proxy, the \pr~algorithm, to assess academic reputation through
the analysis of citation patterns among universities. For the analysis we have selected five different Web of Science categories, corresponding to research subjects studied by well established international academic classifications.

To support the soundness of our work, we have supplied compelling evidence about the close connection of the results of the \pr~algorithm with the scores on two rankings that handle academic reputation in two distinct ways: scores partially based on academic surveys, QS,  or driven by publication and staff figures, ARWU-GRAS. These two academic rankings do have their shortcomings: for instance, ARWU suffers limitations due to aggregation methodologies and with the chosen criteria \citep{Billaut2010}, while the
return rate  of the reputational surveys and reliability of the  statistical data were questioned for QS \citep{Huang2012, Bookstein2010}. However, although those
two ranking methodologies do not enjoy the favour of the whole academic community, we believe they -- at least superficially -- capture some
features of academic excellence. Because in this work we are proposing a new framework for ranking academic institutions,
we therefore felt necessary to compare our results with those two standards, despite the controversy they both stir.

The fact that the \pr~algorithm operates with hard data obtained through the analysis of citation patterns among papers published in peer review publications,  well rooted, therefore,  in a sound and credible mechanism for recognising reputation among researchers and institutions, makes \pr~a very reasonable candidate to be used as a direct mean to assess academic reputation. We believe that the results from our paper   provide a solid argument in favor  of using \pr~scores based only on bibliometric instead of (or along with) estimations  through surveys to measure academic reputation.

The current study offers a new glimpse into the analysis of scientific reputation via \pr~that adds to the large body of literature in which network analysis and bibliometrics are combined. The work carried out in this study may be extended along several research lines,
by exploiting a few Network Science techniques. In this sense, a few interesting research venues
we are planning to explore are for instance related to the measurement of the `boost' in Academic Prestige due to
institutional self--citations (which is expected to enhance the \pr~score and, thus, to correlate positively with
academic ranking scores), to the detection of communities ({\em i.e.} of groups of academic institutions
cross-citing each other more frequently than expected) and, finally, to the relation of the above two properties to the geographical
location of the analysed institutions.

\section*{Acknowledgments}
The authors would like to thank Clarivate Analytics for kindly providing all
the information needed to replicate the results of the PageRank algorithm in the five WoS categories selected for this project. The authors would also like to thank Solange Chavel, Enric Fuster and Sebastian Stride for useful
comments and suggestions.
FAM is supported by the Spanish MINECO grant PTQ-14-06718 of the Torres Quevedo programme. DD is supported by the European Regional Development
Fund (ERDF) and the Galician Regional Government under agreement for funding the Atlantic Research
Center for Information and Communication Technologies (atlanTTIC).

\newpage
\noindent {\bfseries\large References}

\bibliographystyle{spbasic}

\end{document}